# Engineering Dirac states in graphene: Coexisting type-I and type-II Floquet-Dirac fermions

Hang Liu,[1,3] Jia-Tao Sun,[1,3,*] and Sheng Meng[1,2,3,†]

[1] *Beijing National Laboratory for Condensed Matter Physics and Institute of Physics, Chinese Academy of Sciences, Beijing 100190, P. R. China*

[2] *Collaborative innovation center of quantum matter, Beijing 100190, P. R. China*

[3] *University of Chinese Academy of Sciences, Beijing 100049, P. R. China*

**ABSTRACT:** The coupling of monochromatic light fields and solids introduces nonequilibrium Floquet states, offering opportunities to create and explore new topological phenomena. Using combined first-principles and Floquet analysis we show that one can freely engineer Floquet-Dirac fermions (FDFs) in graphene by tuning the frequency and intensity of linearly polarized light. Not only type-II FDFs are created, but they also coexist with type-I FDFs near the Fermi level. Intriguingly, novel topologically nontrivial edge states connecting type-I and type-II Floquet-Dirac points emerge in photodriven graphene, providing an ideal channel to realize electron transport between the two types of Dirac states. Simulating time- and angle-resolved photoelectron spectroscopy suggests that the new coexisting state of type-I and type-II fermions is experimentally accessible. This work implies that a rich FDF phenomenon can be engineered in atomically thin graphene, hinting for developments of novel optoelectronic and quantum computing devices.

Topological semimetals, including Weyl, Dirac, nodal-ring fermions, exhibit extraordinary quantum properties, such as chiral anomaly and quantized transport [1]. Depending on the tilt of the cone around a nodal point, Weyl/Dirac fermion can be classified as type-I and type-II [2-4]. Due to the excessive tilt, type-II fermion possesses a Fermi surface geometry with hole and electron pockets, which is topologically distinct from point-like Fermi surface for type-I fermion. Exotic properties originating from type-II Weyl/Dirac fermions include Fermi arc surface states, large unsaturated magnetoresistance, and Landau level spectrum squeezing [2-18]. Type-I fermions have been found in simple elemental materials of graphene and borophene [19, 20], while the candidates with type-II fermions are only discovered in relatively complex materials, including three-dimensional $WTe_2$, $MoTe_2$, $PtSe_2$ [2, 4, 5, 7, 10, 11, 17, 21, 22], and two-dimensional oxides [23, 24]. The discovery of type-II fermions in common materials such as graphene is highly desirable, which may enable experimentally accessible applications of type-II fermions.

Different from rich topological states with one particular type of fermions, coexistence of distinct classes of topological fermions in a single material has rarely been proposed. Only two cases have been reported: i) coexisting triply degenerate nodal point and Weyl point in ZrTe [25]; ii) coexisting triply degenerate nodal point and nodal ring in TaS and $XB_2$ (X = Sc, Ti, V, Zr, Hf, Nb, Ta) [26, 27]. Due to distinct intriguing properties of type-I and type-II fermions, their coexistence has the potential to produce new topological phenomena, which remains to be explored in real materials. It would be more exciting to create this new coexisting state in graphene, which is significant for both deepening our understanding of band topology and expanding already diverse applications in such a popular simple material.

In contrast to highly challenging conventional approaches to engineer Dirac fermions in graphene, such as large strain [28, 29] and molecular adsorption [30], the laser-driving approach is more effective, producing well-known Floquet-Bloch states. Using the two-dimensional Dirac model confined around K point, Floquet topological insulator, multiple type-I Floquet-Dirac fermions (FDFs) and laser-controlled Fermi velocity of type-I cone are proposed [31-40]. Time-periodic field driven Harper model in three dimensions supports multiple Weyl cones [41, 42], whose dispersion can also be controlled by the external field [43]. The tight binding models manifest the absence of type-II FDFs in graphene, while the variety of these photoinduced cones indicates that type-II fermions, and even its coexistence with type-I fermions, could be created by laser irradiation. Moreover, these states are promising for experimental observation, owing to the simple band structure of graphene [19, 44-46]. In order to predict experimentally accessible Floquet states beyond toy models employed in previous works [31-39], first-principles calculations should encompass the electronic band structure in the whole Brillouin zone (BZ) [47].

Here we predict that not only the number of created multiple FDFs in graphene can be tuned by linearly polarized laser, it can also induce the coexistence of type-I and type-II FDFs near the Fermi level. To create such a state, strong laser intensity is required with increasing photon energy, originating from optical (dynamical) Stark effect. Moreover, novel edge states show up to connect type-I and type-II nodal point in laser-driven graphene, providing the switch between two types of distinct fermions. The predicted Floquet-Dirac states are readily accessible in a well-designed pump-probe setup. This work provides a way to control various types of emergent Floquet-Dirac fermions in the simple graphene monolayer, and opens a new window for investigating the interactions between these emergent quasiparticles.

Beyond the low-energy Dirac model in previous works [31-39], Hamiltonian $H(\bm{k})$ spanning the whole BZ of graphene is constructed by projecting Bloch states onto the $p_z$ Wannier orbital of carbon atoms using Wannier90 interfaced with VASP (see Supporting Information for calculation details) [48-51]. A spatially homogeneous monochromatic laser field with a time-periodic vector potential of $\bm{A}(t) = \bm{A}(t+T)$ is introduced via Peierls substitution $H(\bm{k}) \rightarrow H(\bm{k}, t) = H(\bm{k} - \bm{A}(t))$, which governs the dynamics of laser-driven graphene. Using Floquet theorem [52-55], the time-periodic Hamiltonian $H(\bm{k}, t)$ is mapped onto a time-independent Floquet Hamiltonian $H_F(\bm{k})$, which spans the newly created Hilbert space ($R \otimes T$) as the direct product of original electronic Hilbert space and multiphoton dimensions. Floquet Hamiltonian $H_F(\bm{k})$ is truncated at the fifth order ($n = -2, -1, 0, 1, 2$), which gives sufficiently accurate Floquet-Bloch band structures.

In contrast with massive FDFs induced by circularly polarized laser [31, 32], we explore the behavior of massless FDFs driven by linearly polarized laser $\bm{A}(t) = A_0(\sin(\omega t), 0, 0)$ with light polarization along the armchair (*x*) direction [Fig. 1(a)]. Hybridization between original (*n* = 0) and photon-dressed (*n* $\neq$ 0) bands would modify bands significantly, which may lead to novel electronic states. In order to observe modifications of only *n* = 0 bands around Fermi level $\varepsilon_F$, we first choose the photon energy $\hbar\omega$ = 5 eV, which is large enough to push *n* $\neq$ 0 bands far away from $\varepsilon_F$. The laser amplitude of $A_0$ = 624 V/*c* (corresponding to 1.58 V/Å or 3.32 × 10$^{13}$ W/cm$^2$, here *c* is the velocity of light) is chosen as a start.

The stability of this laser-driven system is estimated, which is crucial for experimental investigation. Floquet states can be observed by irradiating a laser pulse [45, 46, 56], resulted from the good approximation of a long pulse to a continuous wave; this is evidenced from our simulation of angel-resolved photoelectron spectrum. Laser pulse with trapezoidal envelope having 5 cycles (~ 4 fs here) is long enough to produce well defined signatures [56], indicating that not only laser intensity is below graphene damage threshold [57], but the pulse duration is so brief that phonons cannot be excited [58]. Therefore, stable graphene would survive in such a laser irradiation, producing corresponding Floquet-Bloch states.

As shown in Fig. 2(a), under the illumination of above time-periodic laser field, the upper (lower) band at the Fermi level moves down (up), with the most drastic change occurring at M point. As a result, two Dirac cones approach each other along the K-M-K' direction. Band hybridization becomes stronger with increasing laser intensity, which leads to dramatic changes in electronic states. When laser amplitude increases to a critical value of 875 V/*c*, upper and lower bands touch each other around $\varepsilon_F$, forming an anisotropic semi-Dirac cone, with linear and quadratic dispersions along $k_x$ and $k_y$ directions, respectively [Fig. 2(b)]. After $A_0$ exceeding 875 V/*c*, bands at M point are inverted, which leads to a new pair of type-I Floquet-Dirac points around M point [Fig. 2(c,d)]. Both newly created and original Dirac cones have a linear dispersion, yielding a high mobility of electrons in laser-driven graphene. Because linearly polarized laser does not break required symmetries for gapless Dirac cones in graphene [28-30, 59], these Floquet-Dirac cones are still stable against perturbations. More importantly, by comparing with FDFs from crossing *n* $\neq$ 0 bands in previous works [33, 34, 36], engineering FDFs with only *n* = 0 bands here is more effective thanks to its high electron population. Increasing laser intensity further, stronger band inversion drives the new Dirac point moving towards the original Dirac point. At 919 V/$c$, new and original Dirac points merge together, forming Floquet-semi-Dirac fermions again (Fig. S2 [48]).

The topological states can also be controlled by tuning laser frequency. For laser frequency $\hbar\omega$ = 6 eV, new Dirac points show up due to band inversion. Interestingly, the newly-formed Dirac points are transformed to type-II, when the laser amplitude reaches 980 V/$c$ [Fig. 3(a,b)], while the original Dirac points remain type-I. The type-II fermion is clearly characterized by its unique Fermi surface (Fig. S3 [48]): i) when $\varepsilon_F$ is at the energy of type-II Dirac points ($\varepsilon_2$), hole and electron pockets touch at the Dirac point; ii) when $\varepsilon_F$ deviates from the energy of Dirac point ($\varepsilon_F = \varepsilon_2 \pm 5$ meV), the hole and electron pockets get away from each other. Therefore, type-I and type-II FDFs coexist near the Fermi level, with a very small energy separation $\varepsilon_1 - \varepsilon_2 = 0.02$ eV. For comparison, in compressed PdTe$_2$, the type-I and type-II Dirac points have a very large energy separation of ~1 eV, thus is not really "coexisting" [60]. Laser-driven graphene is a better platform to explore interactions between type-I and type-II FDFs, such as the possible realization of magnetic-field-induced Klein tunneling [15]. Separating type-I and type-II Dirac fermions by selective quantum scattering, and the analogous Hawking radiation to form correlated electron-hole pairs by tunneling of type-II Dirac fermions can be potentially observed in such a laser-driven graphene layer [61, 62].

To identify whether the coexisting state can be realized using experimentally accessible laser pulses, we have simulated time- and angle-resolved photoelectron spectroscopy (tr-ARPES) of graphene [48, 63-65]. The intensity of photocurrent $I(\mathbf{k}, \varepsilon, \Delta t)$ for wave vector $\mathbf{k}$, binding energy $\varepsilon$, and pump-probe delay time $\Delta t$ reads [63-65],

$$I(\mathbf{k}, \varepsilon, \Delta t) = Im \sum_a \int dt_1 \int dt_2\, s(t_1, \Delta t) s(t_2, \Delta t) e^{i\omega(t_1 - t_2)} \times G_{aa}^<(\mathbf{k}, t_1, t_2), \qquad (1)$$

where $G_{aa}^<(\mathbf{k}, t_1, t_2)$ is the lesser Green function of laser-driven graphene, and $s$ is the envelop function of probe pulse. Pump-probe setup is designed according to the requirement of realizing the desired Floquet states and the resolution of measurement. Pumping pulse $\mathbf{A}(t) = A_0 p(t)(\sin(\omega t), 0, 0)$ has the photon energy $\hbar\omega$ = 6 eV, and trapezoidal envelope $p(t)$ with the plateau amplitude of 980 $V/c$. Probe pulse $\mathbf{A}(t) = A_0 s(t)(\sin(\omega t), 0, 0)$ has the Gaussian envelop with width of 80 fs, whose peak coincides with the middle of plateau of pump pulse [inset in Fig. 3(c)]. To observe the whole Dirac cone, $\varepsilon_F$ is set at 2 eV, which conforms to the electron doping effect for substrate-supported graphene [66, 67].

Figure 3(c) shows the simulated tr-ARPES by the above setup, which clearly shows two distinct types of Dirac cones. The strongest photocurrent intensity locates at two Dirac points, which are almost degenerate in energy. Most importantly, both the location and dispersion of measured Dirac cones match well with the Floquet-Bloch bands. Therefore, the proposed coexisting state can be probed directly in experiments with the designed setup that is currently accessible.

To demonstrate the unique topological property of the coexisting state, edge states of graphene nanoribbon with the zigzag termination has been explored. The topologically nontrivial edge states connecting type-I and type-II Dirac points show up, as shown in Fig. 3(d), which are significantly different from that in equilibrium state [Fig. S1(d) [48]]. This nontrivial edge state represents an interesting new phenomenon, which may facilitate the scattering and conversion between type-I and type-II Dirac quasiparticles in nonequilibrium conditions. With increasing laser intensity, momentum distribution of the edge states reduces from an arc to a point, which originates from merging type-I and type-II Dirac points [Fig. S2(e,f) [48]].

In order to identify the scope of laser parameters required to realize the coexisting state, phase diagram of Floquet-Dirac states in laser-driven graphene is constructed. As shown in Fig. 1(c), the regions of two type-I FDFs (blue area) and trivially gapped phases (olive area) are distinguishable, which are wedged by the phase of four Floquet-Dirac points (orange and red areas). Moreover, the required laser amplitude to realize the state of four FDFs increases with photon energy. In contrast with the phase of four type-I FDFs (orange area), coexistence of type-I and type-II FDFs requires a stronger laser intensity and higher photon energy (red area). However, the photon energy should be smaller than 9.0 eV; otherwise, the state of four FDFs would disappear, let alone the coexistence of two different kinds of fermions [see Fig. S4 for details [48]]. Therefore, to induce coexisting state, laser parameters should locate at upward diagonal of the diagram, with a photon energy of 6 ~ 8 eV and laser amplitude of 950 ~ 1075 V/$c$. Besides, series of critical Floquet states at phase boundaries can also be induced, such as the state of three Dirac points [blue dashed line and Fig. 2(b)], type-I and type-II anisotropic semi-Dirac cones (olive dashed line and Fig. S2 [48]), and type-III Dirac cone (orange line).

The mechanism behind Floquet phase diagram, especially the coexisting type-I and type-II FDFs phase, is worthwhile to be explored, which could provide a new perspective to design novel Floquet states. As sketched in Fig. 4(a), the variation of $n = 0$ bands at M point is $\Delta\varepsilon = |\Delta\varepsilon_{up}| + |\Delta\varepsilon_{dn}|$; and due to band gap of 3.9 eV at M point, the energy separation between $n = 0$ and $n = \pm 1$ bands is $\delta = \hbar\omega - 3.9$ eV. As the photon energy increases from 3.9 to 6.0 eV with a small $A_0$, evolution of $\Delta\varepsilon$ with laser amplitude deviates from linearity gradually; while with large $A_0$, $\Delta\varepsilon$ always keeps linearity [Fig. 4(b)]. The changes of $\Delta\varepsilon$ conform to $\Delta\varepsilon = \sqrt{A_0^2|M|^2 + \delta^2} - \delta$ with the dipole matrix element $M$ between two states, which indicates band modification is contributed by optical Stark effect [56]. Obviously, in order to obtain four FDFs state through band inversion, the optical Stark effect should be strong enough as $\Delta\varepsilon > 3.9$ eV. Because $\Delta\varepsilon$ decreases with increasing $\hbar\omega$ (or $\delta$), larger $A_0$ is required to realize the transition from two FDFs to four FDFs, which explains wedged shape of four FDFs phase in Fig. 1(c).

Through comparing band evolution in two regions of laser frequency, physical origin of type-II FDFs has been identified. Due to asymmetry of upper and lower bands, just at their inversion, band bending direction around M point have two combinations. For laser with a photon energy of 3.9 ~ 5.7 eV, optical Stark effect is so strong that the bending direction of both upper and lower $n = 0$ bands would be reversed, which leads to type-I FDF [Fig. S5(a) [48]]. In contrast, for laser with the photon energy of 5.7 ~ 9.0 eV, bending direction of only upper band is reversed, which leads the same slope of two bands. Band inversion leads to type-II FDFs, as sketched in Fig. S5(b) [48]. Therefore, both asymmetry of bands in ground state and reasonable strength of optical Stark effect are crucial to realize type-II FDFs, where electron-electron interactions do not have contributions. Therefore, tuning optical Stark effect is a flexible and effective way to realize new matter states in graphene. This is quite different from realizing type-II Weyl cones in continuously driven off-diagonal Harper model, which originates from the crossing between $n = 0$ and $n \neq 0$ bands and the strong quantum effect of driving field [43].

In conclusion, based on first-principles calculations and Floquet theory analysis, new electronic states, including four FDFs and coexisting type-I and type-II FDFs near the Fermi level, is predicted in graphene driven by linearly polarized laser. Moreover, optical Stark effect contributes to formation of the new states, which supports topologically protected Fermi arc edge states connecting the original and emergent Dirac points. Furthermore, the coexisting state can be observed directly by tr-ARPES experiments with designed setups. Graphene provides an effective platform to create and engineer various kinds of FDFs, which are beneficial to understand fundamental interactions between these emergent quasiparticles, such as separation and conversion between type-I and type-II Dirac fermions.

This work was financially supported by the National Key Research and Development Program of China (Grants No. 2016YFA0202300, No. 2016YFA0300902), National Basic Research Program of China (Grant No. 2013CBA01600), "Strategic Priority Research Program (B)" of Chinese Academy of Sciences (Grants No. XDB07030100, No. XDB30000000). We thank Tianjin Supercomputing Center for providing the computing resources.

\* jtsun@iphy.ac.cn

† smeng@iphy.ac.cn

[1] X. C. Huang *et al.*, Observation of the Chiral-Anomaly-Induced Negative Magnetoresistance in 3D Weyl Semimetal TaAs. *Phys. Rev. X* **5**, 031023 (2015).

[2] A. A. Soluyanov *et al.*, Type-II Weyl semimetals. *Nature* **527**, 495-498 (2015).

[3] L. Huang *et al.*, Spectroscopic evidence for a type II Weyl semimetallic state in MoTe2. *Nat. Mater.* **15**, 1155-1160 (2016).

[4] H. Huang, S. Zhou, W. Duan, Type-II Dirac fermions in the PtSe2 class of transition metal dichalcogenides. *Phys. Rev. B* **94**, 121117 (2016).

[5] F. Y. Bruno *et al.*, Observation of large topologically trivial Fermi arcs in the candidate type-II Weyl semimetal WTe2. *Phys. Rev. B* **94**, 121112 (2016).

[6] I. Belopolski *et al.*, Fermi arc electronic structure and Chern numbers in the type-II Weyl semimetal candidate MoxW1−xTe2. *Phys. Rev. B* **94**, 085127 (2016).

[7] M. S. Bahramy *et al.*, Ubiquitous formation of type-II bulk Dirac cones and topological surface states from a single orbital manifold in transition-metal dichalcogenides. arXiv: 1702.08177.

[8] J. Jiang *et al.*, Signature of type-II Weyl semimetal phase in MoTe2. *Nat. Commun.* **8**, 13973 (2017).

[9] K. Koepernik *et al.*, TaIrTe4: A ternary type-II Weyl semimetal. *Phys. Rev. B* **93**, 201101 (2016).

[10] J. Sánchez-Barriga *et al.*, Surface Fermi arc connectivity in the type-II Weyl semimetal candidate WTe2. *Phys. Rev. B* **94**, 161401 (2016).

[11] A. Tamai *et al.*, Fermi Arcs and Their Topological Character in the Candidate Type-II Weyl Semimetal MoTe2. *Phys. Rev. X* **6**, 031021 (2016).

[12] C. Wang *et al.*, Observation of Fermi arc and its connection with bulk states in the candidate type-II Weyl semimetal WTe2. *Phys. Rev. B* **94**, 241119 (2016).

[13] Z. M. Yu, Y. Yao, S. A. Yang, Predicted Unusual Magnetoresponse in Type-II Weyl Semimetals. *Phys. Rev. Lett.* **117**, 077202 (2016).

[14] D. Chen *et al.*, Magnetotransport properties of the type-II Weyl semimetal candidate Ta3S2. *Phys. Rev. B* **94**, 174411 (2016).

[15] T. E. O'Brien, M. Diez, C. W. J. Beenakker, Magnetic Breakdown and Klein Tunneling in a Type-II Weyl Semimetal. *Phys. Rev. Lett.* **116**, 236401 (2016).

[16] S. Tchoumakov, M. Civelli, M. O. Goerbig, Magnetic-Field-Induced Relativistic Properties in Type-I and Type-II Weyl Semimetals. *Phys. Rev. Lett.* **117**, 086402 (2016).

[17] Y. Wang *et al.*, Gate-tunable negative longitudinal magnetoresistance in the predicted type-II Weyl semimetal WTe2. *Nat. Commun.* **7**, 13142 (2016).

[18] F. C. Chen *et al.*, Extremely large magnetoresistance in the type-II Weyl semimetal MoTe2. *Phys. Rev. B* **94**, 235154 (2016).

[19] K. S. Novoselov *et al.*, Electric Field Effect in Atomically Thin Carbon Films. *Science* **306**, 666-669 (2004).

[20] B. Feng *et al.*, Dirac Fermions in Borophene. *Phys. Rev. Lett.* **118**, 096401 (2017).

[21] Z. Wang *et al.*, MoTe2: A Type-II Weyl Topological Metal. *Phys. Rev. Lett.* **117**, 056805 (2016).

[22] Y. Wu *et al.*, Observation of Fermi arcs in the type-II Weyl semimetal candidate WTe2. *Phys. Rev. B* **94**, 121113 (2016).

[23] L. L. Tao, E. Y. Tsymbal, Two-dimensional type-II Dirac fermions in a LaAlO3/LaNiO3/LaAlO3 quantum well. *Phys. Rev. B* **98**, 121102(R) (2018).

[24] M. Horio *et al.*, Two-dimensional type-II Dirac fermions in layered oxides. *Nat. Commun.* **9**, 3252 (2018).

[25] H. Weng, C. Fang, Z. Fang, X. Dai, Coexistence of Weyl fermion and massless triply degenerate nodal points. *Phys. Rev. B* **94**, 165201 (2016).

[26] J.-P. Sun, D. Zhang, K. Chang, Coexistence of topological nodal lines, Weyl points, and triply degenerate points in TaS. *Phys. Rev. B* **96**, 045121 (2017).

[27] X. Zhang, Z.-M. Yu, X.-L. Sheng, H. Y. Yang, S. A. Yang, Coexistence of four-band nodal rings and triply degenerate nodal points in centrosymmetric metal diborides. *Phys. Rev. B* **95**, 235116 (2017).

[28] V. M. Pereira, A. H. Castro Neto, N. M. R. Peres, Tight-binding approach to uniaxial strain in graphene. *Phys. Rev. B* **80**, 045401 (2009).

[29] M. Oliva-Leyva, G. G. Naumis, Understanding electron behavior in strained graphene as a reciprocal space distortion. *Phys. Rev. B* **88**, 085430 (2013).

[30] M. Dvorak, Z. G. Wu, Dirac point movement and topological phase transition in patterned graphene. *Nanoscale* **7**, 3645-3650 (2015).

[31] T. Oka, H. Aoki, Photovoltaic Hall effect in graphene. *Phys. Rev. B* **79**, 081406 (2009).

[32] T. Kitagawa, T. Oka, A. Brataas, L. Fu, E. Demler, Transport properties of nonequilibrium systems under the application of light: Photoinduced quantum Hall insulators without Landau levels. *Phys. Rev. B* **84**, 235108 (2011).

[33] P. M. Perez-Piskunow, G. Usaj, C. A. Balseiro, L. E. F. Foa Torres, Floquet chiral edge states in graphene. *Phys. Rev. B* **89**, 121401 (2014).

[34] G. Usaj, P. M. Perez-Piskunow, L. E. F. Foa Torres, C. A. Balseiro, Irradiated graphene as a tunable Floquet topological insulator. *Phys. Rev. B* **90**, 115423 (2014).

[35] M. Puviani, F. Manghi, A. Bertoni, Dynamics and control of edge states in laser-driven graphene nanoribbons. *Phys. Rev. B* **95**, 235430 (2017).

[36] J.-Y. Zou, B.-G. Liu, Floquet Weyl fermions in three-dimensional stacked graphene systems irradiated by circularly polarized light. *Phys. Rev. B* **93**, 205435 (2016).

[37] K. Kristinsson, O. V. Kibis, S. Morina, I. A. Shelykh, Control of electronic transport in graphene by electromagnetic dressing. *Sci. Rep.* **6**, 20082 (2016).

[38] M. Yang, Z.-J. Cai, R.-Q. Wang, Y.-K. Bai, Topologically trivial and nontrivial edge bands in graphene induced by irradiation. *Physics Letters A* **380**, 2836-2841 (2016).

[39] S. V. Syzranov, Y. I. Rodionov, K. I. Kugel, F. Nori, Strongly anisotropic Dirac quasiparticles in irradiated graphene. *Phys. Rev. B* **88**, 241112(R) (2013).

[40] D. Sticlet, F. Piéchon, Distant-neighbor hopping in graphene and Haldane models. *Phys. Rev. B* **87**, 115402 (2013).

[41] D. Y. H. Ho, J. Gong, Topological effects in chiral symmetric driven systems. *Phys. Rev. B* **90**, 195419 (2014).

[42] L. Zhou, H. Wang, D. Y. H. Ho, J. Gong, Aspects of Floquet bands and topological phase transitions in a continuously driven superlattice. *THE EUROPEAN PHYSICAL JOURNAL B* **87**, 204 (2014).

[43] R. W. Bomantara, J. Gong, Generating controllable type-II Weyl points via periodic driving. *Phys. Rev. B* **94**, 235447 (2016).

[44] Y. W. Son, M. L. Cohen, S. G. Louie, Half-metallic graphene nanoribbons. *Nature* **444**, 347-349 (2006).

[45] Y. H. Wang, H. Steinberg, P. Jarillo-Herrero, N. Gedik, Observation of Floquet-Bloch states on the surface of a topological insulator. *Science* **342**, 453-457 (2013).

[46] F. Mahmood *et al.*, Selective scattering between Floquet–Bloch and Volkov states in a topological insulator. *Nat. Phys.* **12**, 306-310 (2016).

[47] H. Liu, J.-T. Sun, C. Cheng, F. Liu, S. Meng, Photoinduced Nonequilibrium Topological States in Strained Black Phosphorus. *Phys. Rev. Lett.* **120**, 237403 (2018).

[48] *See Supplementary Material for details in calculation method, and other photoinduced Dirac states.*

[49] G. Kresse, J. Furthmuller, Efficient iterative schemes for ab initio total-energy calculations using a plane-wave basis set. *Phys. Rev. B* **54**, 11169-11186 (1996).

[50] J. P. Perdew, K. Burke, M. Ernzerhof, Generalized Gradient Approximation Made Simple. *Phys. Rev. Lett.* **77**, 3865-3868 (1996).

[51] A. A. Mostofi *et al.*, An updated version of wannier90: A tool for obtaining maximally-localised Wannier functions. *Comput. Phys. Commun.* **185**, 2309-2310 (2014).

[52] M. Bukov, L. D'Alessio, A. Polkovnikov, Universal high-frequency behavior of periodically driven systems: from dynamical stabilization to Floquet engineering. *Advances in Physics* **64**, 139-226 (2015).

[53] J. H. Shirley, Solution of the Schrödinger Equation with a Hamiltonian Periodic in Time. *Phys. Rev.* **138**, B979-B987 (1965).

[54] H. Sambe, Steady States and Quasienergies of a Quantum-Mechanical System in an Oscillating Field. *Phys. Rev. A* **7**, 2203-2213 (1973).

[55] K. F. Milfeld, R. E. Wyatt, Study, extension, and application of Floquet theory for quantum molecular systems in an oscillating field. *Phys. Rev. A* **27**, 72-94 (1983).

[56] U. De Giovannini, H. Hubener, A. Rubio, Monitoring Electron-Photon Dressing in WSe2. *Nano Lett.* **16**, 7993-7998 (2016).

[57] A. Roberts *et al.*, Response of graphene to femtosecond high-intensity laser irradiation. *Appl. Phy. Lett.* **99**, 051912 (2011).

[58] S. K. Sundaram, E. Mazur, Inducing and probing non-thermal transitions in semiconductors using femtosecond laser pulses. *Nat. Mater.* **1**, 217 (2002).

[59] A. H. Castro Neto, F. Guinea, N. M. R. Peres, K. S. Novoselov, A. K. Geim, The electronic properties of graphene. *Rev. Mod. Phys.* **81**, 109-162 (2009).

[60] R. C. Xiao *et al.*, Manipulation of type-I and type-II Dirac points in PdTe2 superconductor by external pressure. *Phys. Rev. B* **96**, 075101 (2017).

[61] G. E. Volovik, K. Zhang, Lifshitz Transitions, Type-II Dirac and Weyl Fermions, Event Horizon and All That. *J. Low Temp. Phys.* **189**, 276-299 (2017).

[62] H. Huang, K.-H. Jin, F. Liu, Black hole horizon in the Dirac semimetal Zn2In2S5. *Phys. Rev. B* **98**, 121110(R) (2018).

[63] J. K. Freericks, H. R. Krishnamurthy, T. Pruschke, Theoretical description of time-resolved photoemission spectroscopy: application to pump-probe experiments. *Phys. Rev. Lett.* **102**, 136401 (2009).

[64] M. A. Sentef *et al.*, Theory of Floquet band formation and local pseudospin textures in pump-probe photoemission of graphene. *Nat. Commun.* **6**, 7047 (2015).

[65] A. Farrell, A. Arsenault, T. Pereg-Barnea, Dirac cones, Floquet side bands, and theory of time-resolved angle-resolved photoemission. *Phys. Rev. B* **94**, 155304 (2016).

[66] S. Kim, J. Ihm, H. J. Choi, Y. W. Son, Origin of anomalous electronic structures of epitaxial graphene on silicon carbide. *Phys. Rev. Lett.* **100**, 176802 (2008).

[67] A. Mattausch, O. Pankratov, Ab initio study of graphene on SiC. *Phys. Rev. Lett.* **99**, 076802 (2007).

**Figures**

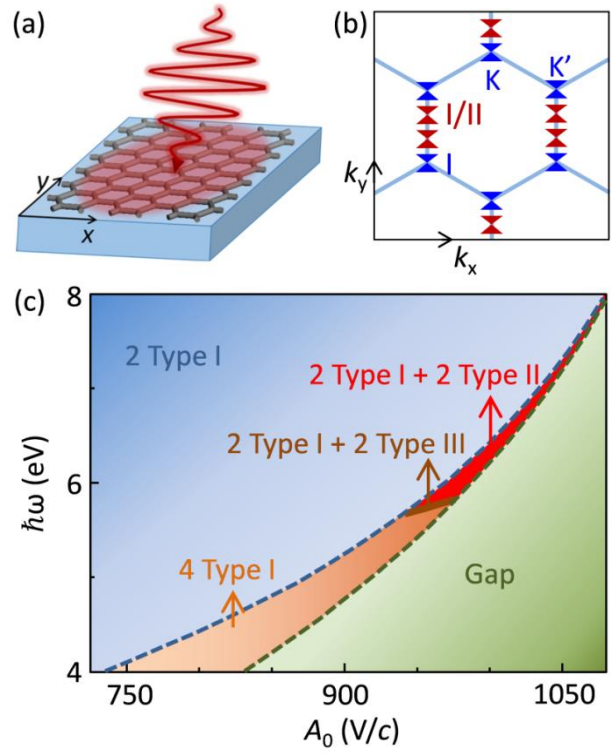

FIG. 1. Multiple Floquet-Dirac points in graphene driven by linearly polarized laser $A(t) = A_0(\sin(\omega t),0,0)$. (a) The laser is polarized along armchair ($x$) direction. (b) Besides original type-I Dirac cones (blue color), new Dirac cones (red color) along KK' path are induced by laser. The photoinduced Dirac cones can be type-I or type-II. (c) Phase diagram of laser-driven graphene. The blue, orange, red and olive zones represent two type-I FDFs, four type-I FDFs, two type-I and two type-II FDFs coexisting, gapped states respectively.

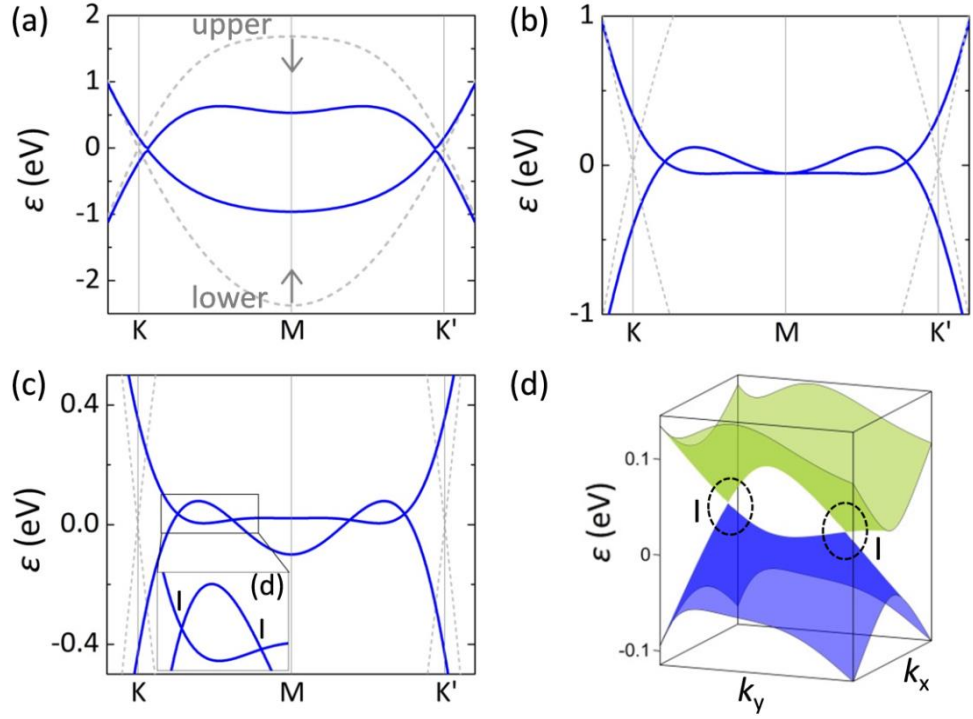

FIG. 2. Floquet-Bloch band structure induced by laser with photon energy $\hbar\omega = 5$ eV and amplitudes of 624 V/$c$ (a), 875 V/$c$ (b), 900 V/$c$ (c,d). Dashed gray lines represent the equilibirum states of graphene.

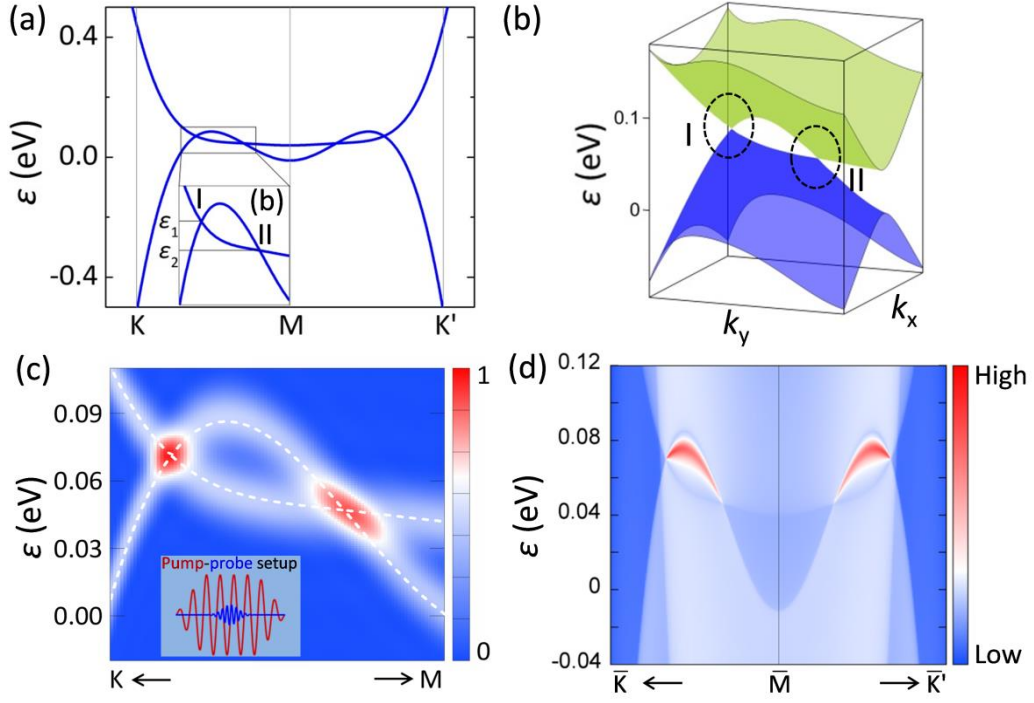

FIG. 3. (a,b) The type-II FDFs are induced by laser with photon energy $\hbar\omega = 6$ eV and amplitude of 980 V/c. (c) The spectrum and white dashed line show the tr-ARPES and Floquet-Bloch bands in (a) respectively. The pump pulse has trapezoidal envelope with photon energy $\hbar\omega = 6$ eV and peak amplitude 980 V/c, while the peak of probe pulse coincide with pump pulse peak in time. (d) Edge states of type-I and type-II FDFs coexisting state.

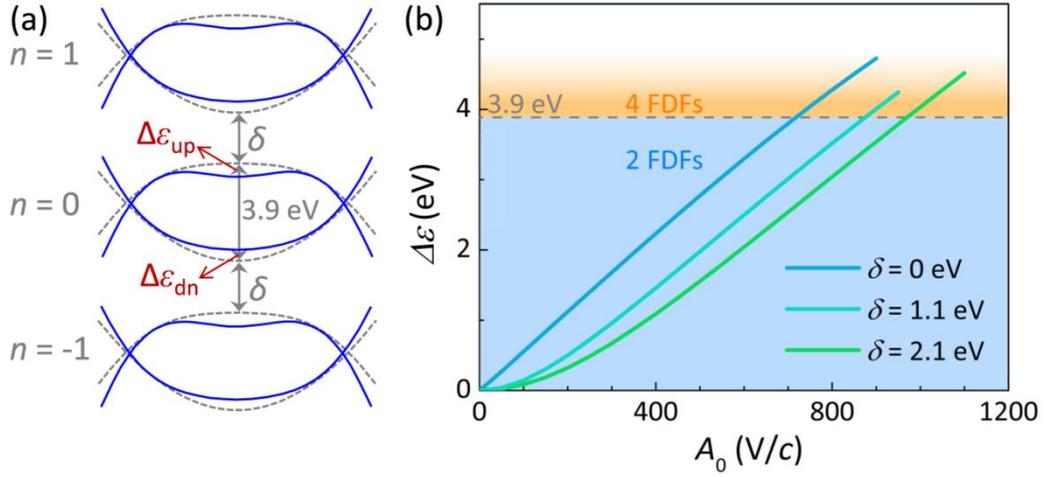

FIG. 4. The shift of quasienergy at M point. (a) Side bands form by absorbing ($n = 1$) and emitting ($n = -1$) one photon. Energy shift of upper and lower bands at M point are labeled as $\Delta\varepsilon_{up}$ and $\Delta\varepsilon_{dn}$ respectively. (b) Total energy shift $\Delta\varepsilon = |\Delta\varepsilon_{up}| + |\Delta\varepsilon_{dn}|$ under laser field with photon energy $\hbar\omega = 3.9, 5.0$ and $6.0$ eV.